%% file: main.tex
\documentclass[letterpaper,twocolumn,10pt]{article}
\usepackage{usenix}

\usepackage{tikz}
\usepackage{amsmath}
\usepackage{amssymb}
\usepackage{booktabs}

\usepackage{multirow}
\usepackage{multicol}
\usepackage{array}    
\usepackage{framed}
\usepackage{hyperref}
\usepackage[capitalise,noabbrev]{cleveref}
\usepackage{listings}
\usepackage{xspace}
\usepackage{csquotes}
\usepackage{dblfloatfix}
\usepackage{placeins}
\usepackage{soul}
\usepackage{microtype}
\usepackage{subcaption}
\usepackage{etoolbox}
\usepackage{colortbl}
\usepackage{makecell}
\usepackage{pifont}

\definecolor{color-a}{RGB}{244, 241, 222}
\definecolor{color-b}{RGB}{129, 178, 154}
\definecolor{color-c}{RGB}{61, 64, 91}
\definecolor{color-d}{RGB}{242, 204, 143}
\definecolor{color-e}{RGB}{224, 122, 95}
\definecolor{color-f}{RGB}{201, 228, 202}
\definecolor{color-g}{RGB}{254, 217, 183}
\definecolor{gray}{gray}{0.9}
\definecolor{light-gray}{gray}{0.4}

\newcommand{\cleanvul}{\textsc{CleanVul}\xspace}
\newcommand{\primevul}{\textsc{PrimeVul}\xspace}
\newcommand{\titanvul}{\textsc{TitanVul}\xspace}

\newcommand{\tool}{\textsc{CPRVul}\xspace}

\sethlcolor{cyan}

\renewcommand{\paragraph}[1]{\vspace{6pt}\noindent{\bf #1}\hspace{8pt}}

\newcommand\realnumberstyle[1]{}
\makeatletter
\newcommand{\linecolor}[3]{
    {\realnumberstyle{#3}}
    \begingroup
    \lst@basicstyle
    \ifnum\value{lstnumber}=#1
        \color{#2}
    \else
        \color{white}
    \fi
    \rlap{\hspace*{\lst@numbersep}
    \color@block{\linewidth}{\ht\strutbox}{\dp\strutbox}
    }
    \endgroup
}

\makeatletter
\newcommand{\storelinecolor}[2]{%
    \expandafter\gdef\csname linecolor@#1\endcsname{#2}%
}

\newcommand{\clearlinecolors}{%
    \@for\next:=1,2,3,4,5,6,7,8,9\do{%
        \expandafter\global\expandafter\let\csname linecolor@\next\endcsname\undefined
    }%
}

\newcommand{\multilinecolor}[1]{%
    {\realnumberstyle{#1}}%
    \begingroup
    \lst@basicstyle
    \ifcsname linecolor@\arabic{lstnumber}\endcsname
        \color{\csname linecolor@\arabic{lstnumber}\endcsname}%
    \else
        \color{white}%
    \fi
    \rlap{\hspace*{\lst@numbersep}%
    \color@block{\linewidth}{\ht\strutbox}{\dp\strutbox}%
    }%
    \endgroup
}

\newcommand{\setlinecolors}[1]{%
    \clearlinecolors
    #1%
}
\makeatother

\newcommand{\linecolorrg}[3]{
    {\realnumberstyle{#3}}
    \begingroup
    \lst@basicstyle
    \ifnum\value{lstnumber}=#1
        \color{#2}
    \else
        \ifnum\value{lstnumber}>#1
            \ifnum\value{lstnumber}<\numexpr#1+\@firstofone#2\relax
                \color{\@secondoftwo#2}
            \else
                \color{white}
            \fi
        \else
            \color{white}
        \fi
    \fi
    \rlap{\hspace*{\lst@numbersep}
    \color@block{\linewidth}{\ht\strutbox}{\dp\strutbox}
    }
    \endgroup
}

\newcommand{\linecolorr}[4]{
    {\realnumberstyle{#4}}
    \begingroup
    \lst@basicstyle
    \ifnum\value{lstnumber}>=#1
        \ifnum\value{lstnumber}<=#2
            \color{#3}
        \else
            \color{red}
        \fi
    \else
        \color{white}
    \fi
    \rlap{\hspace*{\lst@numbersep}
    \color@block{\linewidth}{\ht\strutbox}{\dp\strutbox}
    }
    \endgroup
}

\newcommand{\linecolorrange}[4]{
    {\realnumberstyle{#4}}
    \begingroup
    \lst@basicstyle
    \ifnum\value{lstnumber}>=#1
        \ifnum\value{lstnumber}<=#2
            \color{#3}
        \else
            \color{white}
        \fi
    \else
        \color{white}
    \fi
    \rlap{\hspace*{\lst@numbersep}
    \color@block{\linewidth}{\ht\strutbox}{\dp\strutbox}
    }
    \endgroup
}

\newcommand{\colorlines}[1]{
    \renewcommand{\lst@DefEC}{%
        \lst@CCECUse \lst@ProcessLetter
        \global\let\lst@thestyle\relax
        \edef\tempa{#1}\edef\tempb{\thelstnumber}%
        \ifstrequal{\tempa}{\tempb}%
           {\global\def\lst@thestyle{\color{red}}}{}%
    }
    \expandafter\lst@AddToHook\expandafter{OutputOther}{\lst@thestyle}
}

\lstdefinelanguage{Diff}{
  language=Python,
  sensitive=true,
  morecomment=[f][\color{myred}]-,
  morecomment=[f][\color{mygreen}]+,
}

\begin{document}

\date{}

\title{Beyond Function-Level Analysis: Context-Aware Reasoning for Inter-Procedural Vulnerability Detection}

\author{%
Yikun Li$^{*}$, 
Ting Zhang\textsuperscript{$\ddagger$}, 
Jieke Shi$^{*}$, 
Chengran Yang$^{*}$, 
Junda He$^{*}$, 
Xin Zhou$^{*}$, 
Jinfeng Jiang$^{*}$,\\
Huihui Huang$^{*}$,
Wen Bin Leow$^{\dagger}$, 
Yide Yin$^{\dagger}$, 
Eng Lieh Ouh$^{*}$, 
Lwin Khin Shar$^{*}$, 
David Lo$^{*}$%
\and
$^{*}$Singapore Management University, Singapore\\
$^{\ddagger}$Monash University, Australia\\
$^{\dagger}$GovTech, Singapore
}

\maketitle

\begin{abstract}
Recent progress in machine learning and large language models (LLMs) has improved vulnerability detection, and recent datasets (e.g., \primevul, \titanvul, \cleanvul) have substantially reduced label noise and unrelated code changes. 
However, most existing approaches still operate at the function level, where models are asked to predict whether a single function is vulnerable without inter-procedural context. In practice, vulnerability presence and root cause often depend on callers, callees, and shared global state.
Naively appending such context is not a reliable solution: real-world context is long, redundant, and noisy, and we find that unstructured context frequently degrades the performance of strong fine-tuned code models.

We present \tool, a context-aware vulnerability detection framework that couples \emph{Context Profiling and Selection} with \emph{Structured Reasoning}. 
\tool first clones the full repository for each target function, constructs a code property graph (CPG), and extracts candidate context (callers, callees, and global variables). 
It then uses an LLM to generate security-focused profiles and assign relevance scores, selecting only high-impact contextual elements that fit within the model's context window. 
In the second phase, \tool integrates the target function, the selected context, and auxiliary vulnerability metadata (e.g., commit messages and CVE/CWE information) into a structured prompt to generate reasoning traces, which are used to fine-tune LLMs for reasoning-based vulnerability detection.

We evaluate \tool on three high-quality vulnerability datasets: \primevul, \titanvul, and \cleanvul.
Across all datasets, \tool consistently outperforms function-only baselines, achieving accuracies ranging from 64.94\% to 73.76\%, compared to 56.65\% to 63.68\% for UniXcoder, the strongest fine-tuned baseline.
Specifically, on the challenging \primevul benchmark, \tool achieves 67.78\% accuracy, outperforming prior state-of-the-art approaches, improving accuracy from 55.17\% to 67.78\% (+12.61 points, a 22.9\% relative improvement).
Our ablations further show that neither raw context nor processed context alone benefits strong code models; gains emerge only when processed context is paired with structured reasoning.
Detailed analyses across vulnerability types reveal that \tool's gains are greatest for semantically complex inter-procedural vulnerabilities.
We release context-enriched variants of all three datasets, providing inter-procedural context (19,858 callers, 187,170 callees, and 132,633 global variables) for 23,904 labeled functions across 883 repositories.
\end{abstract}

\section{Introduction}

Machine learning and large language models (LLMs) have driven rapid progress in vulnerability detection, demonstrating promise across a range of benchmarks \cite{marjanov2022machine,lin2020software,perez2021smart,chakraborty2021deep,ding2024vulnerability,lekssays2025llmxcpg,mirsky2023vulchecker,wang2023graphspd}.
However, early vulnerability datasets suffered from quality issues, including noisy labels and unrelated code changes that hindered both training and reliable evaluation \cite{ding2024vulnerability,li2024cleanvul,li2025out,risse2025top}.
In response, substantial effort has been devoted to curating cleaner datasets that address these limitations.
Several recent datasets explicitly address these concerns.
For instance, \primevul~\cite{ding2024vulnerability} extracts vulnerability-fixing commits (VFCs) and retains only those where a single function is modified, thereby filtering out commits with unrelated changes such as test updates, auxiliary bug fixes, or refactorings.
\cleanvul~\cite{li2024cleanvul} and \titanvul~\cite{li2025out} further employ LLM-based filtering to identify and remove noisy or irrelevant code changes, aiming to provide cleaner function-level vulnerability samples. 
In combination, these efforts represent important progress toward more reliable and interpretable vulnerability detection benchmarks.

Despite these improvements, the majority of existing approaches focus on \emph{function-level vulnerability detection}, even when evaluated on these carefully curated datasets \cite{sheng2025llms}.
Models are trained to classify vulnerabilities based solely on the code of an individual function.
While this setting simplifies model input and evaluation, it does not fully reflect the nature of real-world vulnerabilities.
In practice, understanding whether a function is vulnerable often requires reasoning about inter-procedural context, such as the definitions of called functions, how a function's outputs are used by its callers, or how shared global state influences its behavior.
Specifically, improving dataset cleanliness alone does not eliminate this need for context.
Even when unrelated code changes are removed to retain only vulnerability-fixing commits (VFCs) with a single modified function, confirming the presence of a vulnerability in the original function and understanding its root cause may still depend on contextual information outside the function body.
As we demonstrate in \Cref{sec:examples} through real-world examples from \primevul, function-level code alone is often insufficient to identify or explain vulnerabilities.

Incorporating inter-procedural context into vulnerability detection models addresses this limitation but introduces new challenges that remain largely underexplored.
First, contextual information in real codebases is often long, redundant, and noisy~\cite{sintaha2023katana}.
Naively concatenating callers, callees, or global variables can overwhelm the model and, as we show empirically in \Cref{sec:empirical}, does not improve detection performance. 
Second, when both the target function and its surrounding context are non-trivial in size, vulnerability detection becomes a reasoning task rather than a simple classification problem.
Most existing approaches frame vulnerability detection as binary classification~\cite{sheng2025llms} without explicitly guiding models to reason about how contextual elements contribute to vulnerability semantics.
These challenges are coupled with the practical constraints of current LLMs, including limited context windows, performance degradation over long inputs, and the high cost of large-scale fine-tuning.
As a result, there is a lack of study on how to effectively select, structure, and reason over inter-procedural context for vulnerability detection.

\paragraph{Our Solution}
Based on the insight, we propose \tool, a context-aware vulnerability detection framework that couples \emph{context profiling and selection} with \emph{structured reasoning}. 
As illustrated in \Cref{f:overview}, \tool operates in two phases. 
In Phase I \emph{(Context Profiling and Selection)}, \tool clones the full repository corresponding to the target function, constructs a code property graph (CPG), and extracts candidate inter-procedural context, including caller, callee functions, and accessed global variables. It then uses an LLM to generate concise, security-focused profiles for each contextual element and assigns relevance scores with respect to the target function. Based on these scores, \tool selects a compact set of high-impact contextual elements that fit within the LLM's context window. 
In Phase II \emph{(Context-Aware Reasoning and Detection)}, \tool integrates the target function body, the selected contextual elements, and auxiliary vulnerability metadata (e.g., commit messages, Common Vulnerabilities and Exposures (CVE) / Common Weakness Enumeration (CWE) information) into a prompt. The model is prompted to generate explicit structured reasoning traces. These reasoning traces are then used as supervision to fine-tune LLMs, enabling vulnerability detection as a reasoning task grounded in prioritized inter-procedural context.

\paragraph{Evaluation}
We evaluate \tool on three high-quality vulnerability datasets (\primevul~\cite{ding2024vulnerability}, \titanvul~\cite{li2025out}, and \cleanvul~\cite{li2024cleanvul}).
We compare \tool against 
(i) a zero-shot GPT-4.1 baseline, 
(ii) commonly used fine-tuned encoder models (CodeBERT~\cite{feng2020codebert} and UniXcoder~\cite{guo2022unixcoder}), 
(iii) naive context-augmented variants that append raw callers, callees, and global variables (in \Cref{sec:empirical}),
and (iv) representative state-of-the-art vulnerability detection approaches.
Across all datasets, \tool consistently outperforms the strongest function-only baselines, yielding accuracy gains of up to 10.08 points on \titanvul, 11.13 points on \primevul, and 6.70 points on \cleanvul.
On the rigorous \primevul benchmark, \tool achieves 67.78\% accuracy, outperforming state-of-the-art vulnerability detection approaches reported in recent top venues, including USENIX Security'25 LLMxCPG~\cite{lekssays2025llmxcpg}, ICSE'25 Ding et al.'s CoT~\cite{ding2024vulnerability}, and ICSE'26 VulTrial~\cite{widyasari2025let}. 
In particular, \tool improves the best previously reported accuracy from 55.17\% to 67.78\% (+12.61 points, a 22.9\% relative increase).
Our results show that inter-procedural context becomes useful only when models are trained to explicitly reason over it; without reasoning, added context (even after profiling/selection) acts as noise and can hurt performance.
Additional analyses across CWE types, integration strategies, and ablations further validate that \tool's gains arise from the synergy between profiled and selected security-relevant context, and explicitly training models to reason over that context.

\paragraph{Main Contributions} 
Overall, we make the following contributions:

\begin{itemize}
    \item \textbf{Empirical characterization of the context challenge.}
    Using three high-quality datasets, we show that naively appending raw inter-procedural context (callers/callees/globals) does not improve and often degrades the performance of strong fine-tuned code models, highlighting the need for context processing and reasoning.

    \item \textbf{A two-phase framework for context-aware vulnerability detection.}
    We introduce \tool, which uses repository-level static analysis (via CPGs) and LLM-driven profiling/ranking to compress and select inter-procedural context within the context window, and then performs structured reasoning by integrating the target function, selected context, and vulnerability metadata, and fine-tuning LLMs with explicit reasoning traces.

    \item \textbf{Context-enriched benchmarks for inter-procedural vulnerability detection.}
    We construct and release context-augmented variants of \primevul, \titanvul, and \cleanvul (named PrimeVulCTX, TitanVulCTX, and CleanVulCTX) by downloading and analyzing 883 repositories and attaching caller/callee/global definitions to 23,904 labeled functions.
    This process yields a large corpus of inter-procedural context, comprising 19,858 caller functions, 187,170 callee functions, and 132,633 global variables. These datasets provide high-quality benchmarks to support future research on inter-procedural vulnerability detection.

    \item \textbf{Comprehensive evaluation on three datasets with analyses.}
    We evaluate \tool on \primevul, \titanvul, and \cleanvul, showing consistent improvements over strong baselines and prior state-of-the-art, and perform per-CWE analyses, prompt integration comparisons, and ablations to quantify the contribution of each phase.
\end{itemize}

In the spirit of open science, we make our source code and dataset publicly available~\cite{cprvul2026}.

\paragraph{Paper Structure}
The remainder of the paper is organized as follows.
\cref{sec:motivation} presents the background and motivating examples. 
\cref{sec:approach} details the \tool framework, including our context profiling/ranking and reasoning methodology. 
\cref{sec:eva_settings} outlines the evaluation setup, while \cref{sec:eva} presents the experimental results. 
\cref{sec:discussion} discusses the practical implications of our approach. 
\cref{sec:related_work} reviews related work, and \cref{sec:conclusion} concludes the paper and discusses directions for future work.

\section{Background and Motivation}
\label{sec:motivation}

LLMs have recently been explored for vulnerability detection in source code, often operating at the level of individual functions~\cite{sheng2025llms}.
While this setting simplifies input construction and aligns with common vulnerability datasets, many real-world vulnerabilities depend on information that lies outside the target function.
In this section, we discuss the limitations of function-level vulnerability detection and then present empirical evidence showing that naively incorporating context does not effectively address these limitations.

\subsection{Limitations of Function-Level Vulnerability Detection}
\label{sec:examples}

Function-level vulnerability detection assumes that vulnerability-relevant information is contained within the body of the target function~\cite{ding2024vulnerability,li2025out,sheng2025llms}. This assumption simplifies model input and aligns with the design of many existing vulnerability datasets. In particular, datasets such as \primevul are constructed with care to remove vulnerability-fixing commits that modify multiple functions, thereby isolating function-level changes as much as possible \cite{ding2024vulnerability}. 
However, even under this conservative setting, the assumption does not hold for a broad class of real-world vulnerabilities. In practice, confirming whether a function is vulnerable and understanding why often requires contextual information beyond the function body, such as inter-procedural interactions, implicit usage contracts, or execution paths in surrounding code. To illustrate this limitation, we present two real-world vulnerability examples from \primevul. As shown below, despite involving single-function changes, both vulnerabilities depend on contextual information outside the modified function to be correctly identified and understood.

\paragraph{Example I: Condition Pushdown Bug in MariaDB (CVE-2021-46666\footnote{\url{https://github.com/MariaDB/server/commit/2e7891080667c59ac80f788eef4d59d447595772}})}
This is a crash triggered during query optimization in MariaDB when conditions from a \emph{HAVING} clause are pushed into the \emph{WHERE} clause of a grouping view containing aggregate functions with constant arguments (e.g., \emph{SUM(1)}, \emph{MIN(1)}). The root cause lies in how the optimizer reasons about whether an expression is constant.
At the function level, the fix appears very simple. The dataset records the following change to \emph{Item\_direct\_view\_ref::const\_item()}:

\setlinecolors{%
    \storelinecolor{2}{color-g}%
    \storelinecolor{3}{color-f}%
}

\vspace{1mm}
\begin{lstlisting}[language=C, numbers=left, numberstyle=\multilinecolor]
bool const_item() const {
-    return true;
+    return used_tables() == 0;
}
\end{lstlisting}
\vspace{1mm}

Viewed in isolation, this change provides little insight into why it is security-relevant. The vulnerability does not stem from the local logic of \emph{const\_item()} itself, but from how its return value is used by the query optimizer. During condition pushdown, the optimizer relies on \emph{const\_item()} to decide whether predicates can be propagated across query boundaries. An incorrect return value causes non-constant aggregate expressions to be treated as constants, leading to invalid query plans and assertion failures.
Understanding this bug requires context that is unavailable at the function level, including (i) how \emph{const\_item()} is invoked during the pushdown, (ii) the semantics of wrapper classes such as \emph{Item\_direct\_view\_ref}, and (iii) global optimizer state that influences constant-folding decisions. Without this inter-procedural and semantic context, a function-level model cannot determine whether the code above is benign or vulnerable. 

\paragraph{Example II: Invalid Lock Release in USBIP Host Driver (CVE-2018-5814\footnote{\url{https://git.kernel.org/pub/scm/linux/kernel/git/torvalds/linux.git/commit/?id=c171654caa875919be3c533d3518da8be5be966e}})}
This vulnerability is triggered in the Linux \emph{usbip\_host} driver during probe and bind operations, when error paths lead to an invalid lock release. Specifically, under certain execution orders of device bind, unbind, and module loading, the driver attempts to release a spin lock associated with a \emph{bus\_id\_priv} structure that does not exist, resulting in a bad unlock balance and a kernel crash.
At the function level, the fix appears trivial. The dataset records the following change to \emph{put\_busid\_priv()}:

\setlinecolors{%
    \storelinecolor{3}{color-g}%
    \storelinecolor{4}{color-f}%
    \storelinecolor{5}{color-f}%
}

\vspace{1mm}
\begin{lstlisting}[language=C, numbers=left, numberstyle=\multilinecolor]
void put_busid_priv(struct bus_id_priv *bid)
{
-    spin_unlock(&bid->busid_lock);
+    if (bid)
+        spin_unlock(&bid->busid_lock);
}
\end{lstlisting}
\vspace{1mm}

Viewed in isolation, this change appears to be a simple defensive check and offers little indication that the original code is vulnerable.
The original \emph{put\_busid\_priv()} merely releases a spin lock on its argument, with no local evidence that the pointer may be \emph{NULL} or that the lock was not previously acquired, making the implementation appear correct at the function level.
The vulnerability does not stem from the function's local logic, but from how it is invoked along error-handling paths in surrounding driver code, where it may be called with a \emph{NULL} pointer, leading to an invalid lock release. Identifying this issue requires inter-procedural and concurrency context that is unavailable from the function body alone.
These observations suggest that effective vulnerability detection requires reasoning beyond individual function boundaries. At the same time, incorporating additional context introduces new challenges, as discussed next.

\subsection{Empirical Evidence on the Use of Inter-Procedural Context}
\label{sec:empirical}

A common approach to addressing the limitations of function-level detection is to provide models with additional inter-procedural context, such as the definitions of callers, callees, or related global variables.
However, it is unclear whether existing models effectively leverage such context when it is provided without structural organization.
To investigate this, we conduct an empirical study using two frequently used code models: CodeBERT~\cite{feng2020codebert} and UniXcoder~\cite{guo2022unixcoder}. 
These models are established baselines in vulnerability detection and have demonstrated competitive performance on various benchmarks~\cite{jiang2024investigating,wang2024scl,li2024cleanvul,ding2024vulnerability,li2025out}.
The goal of this experiment is to assess whether augmenting function-level inputs with raw context improves detection performance.

\paragraph{Experimental Setup}
We evaluate the models on three high-quality datasets: \primevul, \titanvul, and \cleanvul. We compare two input settings: 
1) Function Body: The model receives only the code of the target function.
2) w/ Raw Context: The target function is concatenated with extracted inter-procedural context, including caller and callee definitions and global variables.
To ensure a fair comparison and account for model architectural constraints, we limit the total input length to 1,024 tokens.
This corresponds to the maximum context window of UniXcoder.
If a function body already exceeds this limit, context cannot be added without causing truncation, making the comparison between function-only and function with context settings invalid.
We prioritize accuracy as the primary evaluation metric.
We do not use F1 scores as the main metric, given that F1 scores can be misleading in balanced datasets.
For example, a model that predicts all samples as positive on a balanced set will achieve a precision of 0.50 and a recall of 1.00, resulting in an F1 score of 0.67.
This score suggests reasonable performance despite the model not being capable of meaningful detection.

\input{tables/motivation}

\paragraph{Results}
\Cref{tb:empirical} presents the performance of CodeBERT and UniXcoder across datasets. 
We notice that providing additional context in an unstructured manner does not improve accuracy. In all tested configurations, the accuracy either remains stagnant or decreases compared to the function-only setting. For example, UniXcoder's accuracy on \primevul drops from 56.65\% to 54.22\% when context is added. Similarly, CodeBERT's accuracy on \cleanvul decreases from 55.67\% to 52.47\%. While F1 scores occasionally increase in the raw context setting, these gains are typically accompanied by a decrease in accuracy, indicating a shift in prediction bias rather than an improvement in vulnerability identification.

\paragraph{Implications}
These results indicate that the challenge lies not in the availability of inter-procedural context, but in how it is integrated into the model. Simply increasing the volume of input information can introduce noise that obscures security-relevant signals. When a model is provided with a large amount of unstructured code, the relevant features of the vulnerability may be lost among irrelevant tokens.

\input{figures/overview}

\subsection{Implications for Reasoning-Based Detection}

Taken together, the case studies and empirical results reveal a critical gap: while inter-procedural context is theoretically necessary, its raw form is practically ineffective.
The MariaDB and USBIP examples demonstrate that distinguishing vulnerability-fixing changes from benign modifications often depends on external factors. 
However, as shown in~\Cref{tb:empirical}, standard fine-tuned models fail to effectively leverage this unstructured context.
These findings suggest that simply providing more code is insufficient; the challenge lies in managing the signal-to-noise ratio. 
Because raw context can overwhelm models with irrelevant tokens, effective detection requires a dual strategy. 
First, the extracted context should be profiled and selected to abstract verbose code into concise, security-focused representations that preserve relevant semantic information.
Second, rather than relying on implicit pattern matching over long sequences, the model should be guided to explicitly reason about how contextual information is used to determine whether the target function is vulnerable.

\section{\tool: Approach}
\label{sec:approach}

\tool is designed to address the limitations of function-level vulnerability detection by enabling structured reasoning over profiled and selected inter-procedural context. Our goal is not merely to provide additional surrounding code to LLMs, but to identify, prioritize, and organize security-relevant contextual information in a way that explicitly supports vulnerability-oriented reasoning under practical input constraints.
As illustrated in \Cref{f:overview}, our approach consists of two main phases.
In Phase I \emph{(Context Profiling and Selection)}, we extract inter-procedural context related to a target function, including its callers, callees, and accessed global variables.
We then apply security-oriented profiling and relevance ranking to select a compact set of high-impact contextual elements that fit within the LLM's context window.
This phase reduces redundancy and noise while preserving information most relevant to vulnerability analysis.
In Phase II \emph{(Context-Aware Reasoning and Detection)}, the target function, the selected contextual elements, and auxiliary vulnerability metadata (e.g., CVE, CWE, and commit information) are provided to the LLM to generate explicit reasoning traces. These reasoning traces are used for supervised fine-tuning, training the model to perform vulnerability detection as a reasoning task grounded in both the function code and its prioritized inter-procedural context, rather than relying solely on local code patterns.

\subsection{Phase I: Context Profiling and Selection}

\paragraph{Context Extraction}
We begin by gathering function-level vulnerabilities in C/C++ from established, high-quality datasets: \primevul~\cite{ding2024vulnerability}, \titanvul~\cite{li2025out}, and \cleanvul~\cite{li2024cleanvul}.
For each vulnerable function, which we refer to as the \emph{target function}, we first download the corresponding source code repository at the vulnerable revision and construct a CPG to enable inter-procedural static analysis. 
Using the CPG, we extract the code context associated with the target function.
The extracted context is composed of three key elements: (1) \emph{Callers}, defined as all functions that invoke the target function; (2) \emph{Callees}, defined as all functions that are invoked by the target function; and (3) \emph{Global Variables} that are accessed or modified within the target function.
Across 883 open-source repositories, our analysis of all target functions yielded a large corpus of contextual data, comprising 19,858 caller functions, 187,170 callee functions, and 132,633 global variables.
This contextual information is crucial, as it provides insights into argument provenance, the behavior of dependent functions, the influence of shared state, and inter-procedural data flows that are not visible from the target function's body alone.

\paragraph{Security Profiling}
While recent LLMs offer expanded context windows, vulnerability detection remains sensitive to input quality rather than input length alone.
Real-world inter-procedural context frequently includes redundant control logic, error handling, and unrelated functionality that can obscure security-relevant cues and hinder effective reasoning, even when token limits are not exceeded.
We therefore introduce a profiling step that compresses extracted context into concise, structured, and security-focused representations, enabling the model to reason over high-signal contextual information rather than raw code volume.
We utilize an LLM to perform \emph{Security Profiling} on each component of the extracted context. For each caller, callee, and global variable, the LLM generates a concise, structured, and security-focused summary.
For example, a caller's profile includes its data origin (e.g., \emph{User Input}, \emph{Network}), data transformations (e.g., \emph{Sanitized}, \emph{Raw/Un-validated}), and usage of the target function's return value. 
Similarly, a callee's profile includes its security risk level and a justification (e.g., ``Performs memory copy without bounds checking''). 
A global variable's profile contains a succinct description of its role and security implications.
This profiling stage transforms verbose code into concise, structured, security-focused metadata, preventing potentially useful information from being treated as noise during fine-tuning.

\input{tables/prompt}

\paragraph{Relevance Ranking}
To further refine the context, we introduce a ranking step to prioritize the most critical information. 
We employ an LLM to assess the security relevance of each profiled context component (callers, callees, and global variables) with respect to the target function. 
The model is prompted to analyze the target function's body alongside the generated security profiles and assign a relevance score to each context component, reflecting its importance for vulnerability analysis.
Based on these scores, we select and retain the most security-relevant context components that fit within the LLM's context window, while discarding less relevant information.
This ranking step serves as a selection mechanism that concentrates the model's attention on a compact set of high-impact contextual elements.
By doing so, it ensures that limited input capacity is devoted to information most likely to influence vulnerability reasoning.

\subsection{Phase II: Context-Aware Reasoning and Detection}

\paragraph{Reasoning Generation and Fine-Tuning}
In the second phase, we perform context-aware vulnerability detection through structured reasoning. 
A key innovation of our approach is the design of the reasoning templates shown in \Cref{tb:prompts}, which provide the LLM with a comprehensive information suite to maximize the quality of generated reasoning traces.
Beyond metadata such as CVE descriptions, CWE classifications, and commit messages, we also provide the LLM with the code diff representing the vulnerability fix.
By including the diff, the LLM can more easily identify the precise modifications that addressed the flaw, enabling it to better understand the core mechanism of the vulnerability and the logic behind the security patch.
This rich set of inputs is combined with the prioritized inter-procedural context (callers, callees, and global variables) identified in Phase I. 
The model is prompted to generate explicit reasoning traces that synthesize these diverse signals to justify whether the target function is vulnerable or benign.
These reasoning traces are then used for supervised fine-tuning, guiding the LLM to better utilize contextual information during vulnerability detection, rather than relying solely on implicit pattern matching in a binary classification setting.
Through fine-tuning, the model learns to jointly leverage the target function and its prioritized, security-focused context, to reason about inter-procedural interactions and implicit security assumptions, and to produce a final vulnerability decision grounded in its reasoning process.
Specifically, at inference, \tool utilizes the target function and processed context as input without additional information.

\section{Evaluation Setup}
\label{sec:eva_settings}

To evaluate the effectiveness of our approach, we design our experiments to answer the following research questions (RQs).

\paragraph{RQ1: Overall Effectiveness}
This research question evaluates the overall detection performance of our approach. We compare against a baseline model that operates solely on the target function without additional contextual information.

\paragraph{RQ2: Performance Across Vulnerability Types}
Aggregate performance metrics may mask differences across vulnerability types. Since vulnerabilities exhibit diverse characteristics (e.g., buffer overflows versus injection flaws), we analyze performance across the most frequent CWEs to understand where our approach is more or less beneficial.

\paragraph{RQ3: Impact of Context Integration Strategies}
The manner in which contextual information is presented to LLMs can influence their behavior. We evaluate three alternative integration strategies: placing context before the target function, after the target function, or embedding it as inline comments. This analysis allows us to assess the sensitivity of performance to input structuring choices.

\paragraph{RQ4: Contribution of Individual Phases}
We conduct an ablation study to isolate the contribution of each phase. By selectively removing each phase and measuring the resulting performance change, we assess their relative importance for accurate vulnerability detection.

\subsection{Datasets}

We evaluate our approach on three high-quality vulnerability datasets: \primevul~\cite{ding2024vulnerability}, \titanvul~\cite{li2025out}, and \cleanvul~\cite{li2024cleanvul}. 
These datasets were selected specifically for their high label validity, defined as the percentage of samples labeled as vulnerable that contain confirmed security vulnerabilities. 
High validity is essential for training reliable detection models and avoiding the noise inherent in uncurated collections~\cite{li2025out}. 
Specifically, \primevul and \cleanvul report validity rates of 86\%~\cite{ding2024vulnerability} and 91\%~\cite{li2024cleanvul}, respectively, based on manual inspection.
\titanvul, which employs multi-agent LLM verification for cleaning, achieves a validity of 92\%.
Consistent with dominant trends in vulnerability research~\cite{fan2020ac,ding2024vulnerability,chen2023diversevul}, we focus our study on C/C++ repositories. 
To enable context-aware analysis and ensure a fair comparison across all models, we apply a length constraint to the data.
This is necessary because real-world functions can be very long; for example, \primevul contains functions exceeding 276,938 tokens, which exceeds the processing capacity of most models. 
Furthermore, as discussed in \Cref{sec:empirical}, the architecture of strong code model baselines such as UniXcoder is restricted by a maximum context window of 1,024 tokens.
To maintain a consistent evaluation environment across all datasets and model types, we filter the data to retain only target functions with a length of up to 1,024 tokens.
This threshold has been adopted in prior studies~\cite{jiang2024investigating,wang2024line,yang2025sparsecoder}, accommodating the architectural limits of encoder models while reserving sufficient space within the context window.
After applying these filters, our experimental datasets consist of 13,038 vulnerable and benign functions from \titanvul, 5,508 from \primevul, and 4,348 from \cleanvul.

\subsection{Model Choices} 

Our framework employs different models for context preparation, reasoning-based detection, and comparative evaluation.
For Phase I (\emph{Security Profiling} and \emph{Relevance Ranking}) and the reasoning generation of training data for Phase II, we utilize GPT-4.1~\cite{openai2025gpt41}.
We select this model for its advanced instruction-following and reasoning capabilities.
For the final detection stage, we fine-tune two variants of the Qwen2.5-Coder family (7B and 32B)~\cite{hui2024qwen2}, which serve as the base LLMs for \tool.
We select Qwen2.5-Coder for three reasons: (i) it is a strong code-specialized LLM, (ii) its open weights enable controlled task-specific fine-tuning, and (iii) it supports instruction-style prompting and supervised training with structured reasoning traces, which aligns with our context-aware detection framework.
For brevity, we refer to these models as Qwen2.5-7B and Qwen2.5-32B in the remainder of the paper.
To rigorously evaluate \tool, we compare it against CodeBERT~\cite{feng2020codebert} and UniXcoder~\cite{guo2022unixcoder}. 
These models serve as widely adopted baselines in the field of vulnerability detection. Recent empirical studies consistently identify UniXcoder as a top performer among code model architectures.
For instance, Jiang et al. report that it achieves the highest accuracy and F1 scores on standard benchmarks like Devign and ReGVD \cite{jiang2024investigating}.
On cleaned datasets such as \primevul, UniXcoder remains a competitive code model baseline~\cite{ding2024vulnerability}. 
These baselines provide a reference point, allowing us to quantify exactly how much \tool improves upon existing technology.

\subsection{Evaluation Metrics}
\label{sec:metrics}

We adopt four standard metrics to evaluate model performance: Accuracy, Precision, Recall, and F1-score. 
Given that all evaluation datasets are balanced with paired vulnerable and benign samples, we consider \emph{Accuracy} to be the primary indicator of performance.
In balanced classification settings, accuracy provides a direct and intuitive measure of a model's effectiveness.
In contrast, metrics like the F1-score can be misleading. For instance, a model that naively classifies all functions as \emph{vulnerable} would achieve a perfect recall of 1.0 but a precision of only 0.5, resulting in an F1-score of 0.67 despite failing to discriminate between vulnerable and benign samples.
Formally, Accuracy is the ratio of correct predictions to the total number of samples: $\text{Accuracy} = \frac{TP + TN}{TP + TN + FP + FN}$.
Precision measures the fraction of predicted vulnerable functions that are truly vulnerable, reflecting the false-alarm rate and the manual triage burden: $\text{Precision} = \frac{TP}{TP + FP}$.
Recall measures the fraction of true vulnerable functions that are correctly identified, reflecting the risk of missing real vulnerabilities: $\text{Recall} = \frac{TP}{TP + FN}$.
Finally, the F1-score is the harmonic mean of Precision and Recall, summarizing the trade-off between false positives and false negatives: $\text{F1-Score} = 2 \times \frac{\text{Precision} \times \text{Recall}}{\text{Precision} + \text{Recall}}$.

\subsection{Implementation Details}

We implement our pipeline using PyTorch and the Hugging Face Transformers library. For the profiling and selection phases, we access the GPT-4.1 model via its standard API. For the detection phase, we fine-tune the Qwen2.5-Coder models (7B and 32B) using Low-Rank Adaptation (LoRA) to efficiently manage memory and computational costs. We configure LoRA with a rank $r=256$ and a scaling factor $\alpha=512$, applying adapters to all linear layers.
Models are fine-tuned for 5 epochs with a learning rate aligned with the model's pre-training configuration to ensure stability. We employ a checkpointing strategy that saves the model with the highest accuracy on the validation set. All experiments are conducted on a high-performance server running Ubuntu 22.04.5 LTS, equipped with NVIDIA H100 GPUs and an Intel(R) Xeon(R) Platinum 8480C CPU.

\section{Evaluation Experiments}
\label{sec:eva}

\subsection{RQ1: Overall Effectiveness}
\label{sec:rq1}

We evaluate the performance of \tool relative to two baseline families shown in \Cref{tb:performance}: (i) a zero-shot GPT-4.1 baseline and (ii) commonly used fine-tuned code models (CodeBERT~\cite{feng2020codebert} and UniXcoder~\cite{guo2022unixcoder}).
In all baseline settings, the model receives only the target function body and outputs a binary vulnerability prediction without intermediate reasoning.
For \tool, we apply security profiling and relevance ranking to select high-priority context, and then perform reasoning-based detection using fine-tuned Qwen2.5-Coder models (7B and 32B).
As defined in \Cref{sec:metrics}, we adopt accuracy as our primary metric due to the balanced nature of all datasets (i.e., \titanvul, \primevul, \cleanvul).
We also report precision, recall, and F1 to characterize error trade-offs.

\input{tables/rq1_performance}

\paragraph{Performance of Baselines}
Consistent with recent literature, function-level baselines exhibit significant limitations. Zero-shot GPT-4.1 achieves near-random accuracy (around 51\%) across all datasets. While it maintains high recall (76.74\% to 83.30\%), its precision is low (around 51\%), indicating a strong bias towards over-flagging code as vulnerable when context is missing. Among fine-tuned strong baseline models, UniXcoder consistently outperforms CodeBERT, establishing itself as the strongest baseline with accuracies ranging from 56.65\% on \primevul to 63.68\% on \titanvul. However, even the strongest baseline struggles to break the 60\% accuracy barrier on the \primevul and \cleanvul benchmarks. These results indicate that conventional binary classification models struggle to learn effectively from high-quality, low-noise datasets, confirming that intra-procedural features alone are insufficient for robust detection in realistic scenarios \cite{ding2024vulnerability}.

\paragraph{Performance of \tool}
\Cref{tb:performance} shows that \tool consistently outperforms all baselines across all datasets. 
With the Qwen2.5-32B backbone, \tool achieves substantial accuracy improvements over the strongest baseline (UniXcoder): on \titanvul, accuracy increases from 63.68\% to 73.76\%; on \primevul, from 56.65\% to 67.78\%; and on \cleanvul, from 58.24\% to 64.94\%. 
Notably, on \primevul and \cleanvul, where baseline performance barely exceeds random guessing on this balanced dataset (56.65\% and 58.24\% respectively), \tool advances accuracy to substantially higher levels (67.78\% and 64.94\%), demonstrating that the combination of profiled inter-procedural context and structured reasoning enables the model to discriminate effectively between benign and vulnerable patterns.
The smaller Qwen2.5-7B model also achieves gains, improving accuracy by 6.00 points on \primevul, 4.78 points on \titanvul, and 3.94 points on \cleanvul, validating that the benefits of context-aware reasoning extend across different model capacities.

\input{tables/rq1_sota}

\paragraph{Comparison with State-of-the-Art Approaches}
We further compare \tool with state-of-the-art vulnerability detection approaches evaluated on \primevul, as shown in \Cref{tb:sota}.
We focus on \primevul because it is a widely adopted benchmark in recent vulnerability detection research, and many state-of-the-art methods report their results on this dataset, enabling fair comparison.
These methods include systems that leverage chain-of-thought prompting, multi-agent LLM analysis, or LLM-assisted program analysis for vulnerability detection, such as Ding et al.'s CoT approach~\cite{ding2024vulnerability}, GPTLens~\cite{hu2023large}, VulTrial~\cite{widyasari2025let}, and the recent USENIX Security'25 work LLMxCPG~\cite{lekssays2025llmxcpg}, which uses LLMs for code slicing and vulnerability detection.
As reported in the table, these methods achieve accuracies ranging from 50.36\% to 55.17\%, with F1 scores ranging from 28.14 to 57.63.
Compared to the zero-shot GPT-4.1 baseline (in \Cref{tb:performance}), GPTLens \cite{hu2023large} and VulTrial \cite{widyasari2025let} improve accuracy, indicating that explicit reasoning or multi-agent analysis can be beneficial over direct zero-shot inference.
Notably, \tool substantially outperforms all reported state-of-the-art methods on \primevul.
\tool achieves 67.78\% accuracy, exceeding the best prior approach by +12.61 percentage points in accuracy (a 22.9\% relative improvement) and +10.30 points in F1 (a 17.9\% relative improvement).
These gains indicate that, while prior work has explored various strategies for vulnerability detection, our approach of combining context profiling and selection with fine-tuned reasoning detection yields larger improvements on \primevul than the SOTA approaches.

\paragraph{Impact on Precision}
The results in \Cref{tb:performance} show that \tool generally improves precision relative to baseline methods, although the magnitude of the improvement varies across datasets and model sizes. On \titanvul, \tool increases precision from 61.99\% with the strongest baseline (UniXcoder) to 72.88\% when using Qwen2.5-32B. A similar pattern is observed on \primevul, where precision rises from 55.34\% (UniXcoder) to 67.62\% under \tool. On \cleanvul, \tool achieves the precision of 66.00\%, improving over the UniXcoder baseline (56.62\%). These results suggest that our approach can help the detector make more selective positive predictions. In contrast, several baselines achieve relatively high recall but lower precision, indicating a tendency to over-predict vulnerabilities. From a practical standpoint, improvements in precision are especially important for vulnerability detection systems, as a high false-positive rate can impose substantial manual triage costs and hinder adoption in real-world development workflows \cite{alahmadi202299,ami2024false}.

\begin{framed}
\noindent \tool outperforms the strongest function-level baselines across all datasets, achieving accuracy gains of \textbf{6.7} to \textbf{11.1} percentage points. On the rigorous \primevul benchmark, \tool establishes a new state-of-the-art, raising the best prior accuracy from \textbf{55.17\%} to \textbf{67.78\%} (\textbf{22.9\%} relative improvement).
\end{framed}

\subsection{RQ2: Performance Across Vulnerability Types}
\label{sec:rq2}

\input{tables/rq2_cwe}

To understand the characteristics and limitations of \tool in practice, we evaluate performance across individual CWE types.
We focus this analysis on \primevul and \titanvul, as we exclude \cleanvul due to the absence of CWE annotations for the majority of its samples.
\Cref{tb:cwe} presents per-CWE results for models fine-tuned from Qwen2.5-32B, comparing our context-aware approach \tool to the best baseline from RQ1. We report accuracy, precision, recall, and F1 for each CWE, along with absolute improvements in accuracy ($\Delta$ Acc) and F1 ($\Delta$ F1).

\paragraph{Overall Improvement Trends}
Across all CWE instances in \Cref{tb:cwe}, \tool achieves positive accuracy improvements in 28 out of 30 cases (93.3\%), with only two CWEs exhibiting accuracy degradation ($\Delta$ Acc $<$ 0), both on \titanvul.
Averaged across datasets, \tool improves accuracy by +10.79 percentage points, with per-dataset averages of +12.72 points on \primevul and +8.85 points on \titanvul.
At the CWE level, accuracy improvements range from -6.90 points (CWE-415 on \titanvul) to +27.09 points (CWE-264 on \primevul).
This substantial variation across vulnerability types suggests that the combination of contextual information and reasoning-based detection provides differential benefits depending on the nature of the vulnerability. 

\paragraph{High-Impact CWEs}
The most substantial accuracy improvements occur for vulnerability types that benefit from both contextual information and reasoning-based detection. On \primevul, CWE-264 (Permissions, Privileges, and Access Controls) shows the largest overall improvement, with accuracy increasing from 52.08\% to 79.17\% (+27.09 points) and F1 from 59.65\% to 79.17\% (+19.52 points).
One possible explanation is that access control vulnerabilities often involve authorization logic distributed across multiple functions, where contextual information can surface the presence or absence of checks in caller functions, and reasoning chains can help relate these checks to the security of the target function.
A similar pattern is observed for CWE-200 (Exposure of Sensitive Information), where accuracy improves from 53.57\% to 76.79\% (+23.22 points) on \primevul.
In such cases, contextual information may help identify sensitive data sources, such as global variables containing credentials or keys, while reasoning chains can assist in relating these sources to potential exposure points.
Resource management vulnerabilities (CWE-399) also show notable improvements.
On \titanvul, accuracy increases from 56.45\% to 71.88\% (+15.43 points), and on \primevul from 50.00\% to 70.00\% (+20.00 points). 
These gains suggest that combining contextual information with reasoning may be helpful for analyzing resource lifecycle patterns, such as matching allocations with corresponding cleanup across function boundaries. 

\paragraph{Low-Impact CWEs}
Certain vulnerability types show minimal or even negative accuracy improvements, revealing limitations of the approach.
Most notably, CWE-415 (Double Free) exhibits substantially degraded accuracy on \titanvul, dropping from 68.97\% to 62.07\% (-6.90 points), with an even more severe F1 degradation of -11.71 points. 
On \primevul, while accuracy increases from 50.00\% to 62.50\% (+12.50 points), the F1 score decreases by -15.79\%.
The relatively weaker performance on double-free vulnerabilities suggests a potential limitation of the current approach.
While contextual information can provide useful structural cues about program organization, detecting double-free issues may additionally require temporal reasoning to track when and under what conditions the same memory object is freed across different execution paths.
The generated reasoning chains attempt to account for aliasing relationships and control-flow conditions, but this analysis may become less reliable as the number of feasible execution paths increases.

\begin{framed}
\noindent \tool achieves positive accuracy improvements on \textbf{93.3\%} of evaluated vulnerability types. The approach is most effective for context-dependent classes like Access Control (CWE-264) and Resource Management (CWE-399), where it yields gains of up to \textbf{+27.09 points}. However, improvements are limited or negative for temporal vulnerabilities (e.g., Double Free), highlighting that \tool's primary strength lies in spatial inter-procedural reasoning rather than temporal state tracking.
\end{framed}

\subsection{RQ3: Impact of Context Integration Strategies}

To answer this RQ, we investigate whether the method of presenting contextual information to the model affects its detection performance. The underlying hypothesis is that the structure of the input prompt can influence an LLM's reasoning process.
We compare three methods for combining the target function's source code with the processed contextual information.
The first strategy, \emph{Insert-Before}, prepends the entire context as a single, contiguous block of text before the target function's code.
The second, \emph{Insert-After}, appends the same context block after the function's code.
The third strategy, \emph{As-Comments}, inserts the contextual information directly into the target function's body as comments. 
These three strategies are also evaluated on the \primevul, \titanvul, and \cleanvul datasets using the Qwen2.5-32B model.

\input{tables/rq3}

\paragraph{Comparison of Integration Strategies} 
\Cref{tb:integration} summarizes the performance of the three context integration strategies across three datasets.
While performance remains relatively stable across strategies, indicating that \tool is not overly sensitive to context placement, a consistent pattern emerges where inserting context after the target function (\emph{Insert-After}) yields the highest accuracy and F1 scores.
On \titanvul, \emph{Insert-After} achieves 73.76\% accuracy, outperforming \emph{Insert-Before} and \emph{As-Comments} by 1.76 and 1.50 percentage points, respectively.
This advantage is more pronounced on \primevul, where \emph{Insert-After} reaches 67.78\% accuracy, exceeding the alternative strategies by up to 3.18 points.
On \cleanvul, \emph{Insert-After} maintains this lead.
While \tool demonstrates robustness to context placement, the \emph{Insert-After} strategy offers a consistent performance edge.
These performance gaps suggest that \tool's effectiveness primarily stems from the selection and structuring of relevant contextual information rather than the specific prompt layout. 
However, the consistent better performance of \emph{Insert-After} indicates that presenting the target function before its context is more favorable for the reasoning process.
This ordering may allow the model to form an initial representation of the function under analysis before incorporating contextual cues as supporting evidence, whereas prepending or interleaving context may introduce distractions.
Importantly, the lack of large performance degradation across all strategies confirms that \tool is robust to variations in context integration.

\begin{framed}
\noindent Different context integration strategies lead to limited performance differences. Across datasets, inserting context after the target function improves accuracy by \textbf{1.5} to \textbf{3.2} points compared to alternative strategies. While the overall impact is modest, \emph{Insert-After} consistently provides the strongest results and is therefore adopted as the default integration strategy in \tool.
\end{framed}

\subsection{RQ4: Contribution of Individual Phases}
\label{sec:ablation}

To assess the contribution of different phases in \tool, we conduct an ablation study comparing the full system with variants that selectively remove Phase I (context profiling and selection) or Phase II (context-aware reasoning and detection). The results are summarized in \Cref{tb:ablation}. All experiments use Qwen2.5-32B to isolate the effect of system components rather than model capacity.

\input{tables/rq4_ab1}

\paragraph{w/o Phase I (Context Profiling and Selection)}
Removing Phase I leads to a consistent degradation in performance across all datasets. 
On \primevul, accuracy decreases from 67.78\% to 65.42\% (-2.36 points), while on \cleanvul and \titanvul the drops are -2.42 and -1.56 points, respectively. 
Although these decreases are smaller than those observed when removing Phase II, they are non-trivial in this setting: even the strongest function-only baselines achieve accuracies only 6.65 points above random guessing (50\%) on \primevul (see \Cref{tb:performance}).
In this context, 2 to 3 points drop corresponds to a substantial fraction of the total improvement margin. 
These results suggest that Phase I's context profiling and selection plays an important supporting role by ensuring that the reasoning component operates on concise, security-relevant representations rather than being exposed to redundant or noisy context.
The consistency of this effect across datasets indicates that context curation is an important factor for effective reasoning-based vulnerability detection.

\paragraph{w/o Phase II (Context-Aware Reasoning and Detection)}
Removing Phase II leads to a pronounced reduction in performance across all datasets.
On \titanvul, accuracy decreases from 73.76\% to 52.38\% (-21.38 points), and F1 decreases from 74.54 to 39.84.
Similar reductions are observed on \primevul (-15.24 accuracy points) and \cleanvul (-12.53 points).
These results suggest that the reasoning and detection stage plays an important role in utilizing the profiled contextual information.
In the absence of Phase II, the model appears less able to convert contextual cues into accurate vulnerability predictions, even when such information is available.

\paragraph{Impact on Precision} 
The ablation results highlight that \tool is particularly effective at improving precision, which is often a primary concern in real-world vulnerability triage \cite{alahmadi202299,ami2024false}. When Phase II (reasoning) is removed, precision collapses significantly, dropping from 72.88\% to 54.33\% on \titanvul and from 66.00\% to 55.68\% on \cleanvul. 
Removing Phase I (profiling) also results in notable precision losses of 4.75 points on \primevul and 3.97 points on \cleanvul. 
These drops demonstrate that the synergy of profiling relevant context and structured reasoning is essential for filtering out false positives, ensuring that the model does not merely flag functions based on the presence of any context but actually understands the security implications.

\input{tables/rq4_ab2}

\paragraph{Function Body vs w/ Phase I}
To assess whether our processed context is beneficial on its own, we also apply Phase I (context profiling and selection) on two strong code model baselines, CodeBERT and UniXcoder, and fine-tune them using the same binary classification objective as the function-only setting. \Cref{tb:empirical_2} reports results on \primevul, \titanvul, and \cleanvul. 
As can be seen, adding Phase I context does not improve accuracy for either model on any dataset and often degrades performance, with the largest drops observed for CodeBERT. This suggests that, even after profiling and selection, conventional encoder classifiers struggle to incorporate inter-procedural signals and can be distracted by additional inputs.
In contrast, the full \tool pipeline achieves gains when the selected context is paired with Phase II structured reasoning, reinforcing that the combination of \emph{structured reasoning} and \emph{context profiling and selection} is key for leveraging inter-procedural information effectively.

\begin{framed}
\noindent Phase I (Profiling and Selection) and Phase II (Reasoning) are complementary and are most effective when used together. Phase I yields modest but consistent improvements when reasoning is present, while processed context alone does not provide measurable benefits for code model baselines. Phase II accounts for a larger share of the observed gains. Overall, \tool's improvements come from the \textbf{synergy} of selecting high-impact context and explicitly training the model to reason over it.
\end{framed}

\vspace{-1mm}
\section{Discussions}
\label{sec:discussion}

\paragraph{Context Matters, but Naive Context Injection Hurts}
A straightforward approach to inter-procedural vulnerability detection is to simply concatenate additional context (e.g., callers, callees, and global variables) with the target function and allowing the model to determine the relevance of those inputs.
Intuitively, this should improve results because many real-world vulnerabilities depend on inter-procedural interactions. 
However, our study in \Cref{sec:empirical} shows the opposite: appending raw context to strong code models (e.g., UniXcoder) consistently fails to improve accuracy and often degrades performance. 
These findings suggest that the primary bottleneck is not the lack of access to context, but the model's limited ability to distill signal from noise under practical context-window constraints. 
In real-world repositories, inter-procedural context is often long, redundant, and only weakly related to the security flaw. 
Naively adding such information can dilute important cues, introduce spurious correlations, and increase the risk that critical evidence is truncated or lost due to attention erosion over longer inputs.

\paragraph{The Necessity of Reasoning for Context-Aware Detection} 
Our results indicate that inter-procedural vulnerability detection is fundamentally a reasoning task rather than a simple pattern-matching problem. 
To achieve high accuracy, a model should synthesize evidence across multiple locations and justify whether specific inter-procedural interactions constitute a security weakness. 
This requirement explains why unstructured context often proves counterproductive for standard binary classifiers.
The empirical evidence in \Cref{tb:empirical,tb:empirical_2} demonstrates that providing either raw or profiled and selected context to strong code models fails to yield improvements when structured reasoning is absent. 
These findings imply that for inter-procedural detection, explicit reasoning is critical to guide the interpretation of contextual elements. 
Without a reasoning framework to link context to the target function's logic, additional information acts as noise that obscures rather than clarifies the vulnerability's root cause.

\vspace{-1mm}
\section{Related Work}
\label{sec:related_work}

\paragraph{Automated Vulnerability Detection}
Most existing approaches formulate vulnerability detection as function-level binary classification~\cite{woo2023v1scan,lekssays2025llmxcpg,sheng2025llms,xiao2020mvp,li2021arbitrar,yin2022finding}.
To improve performance within this setting, prior work has explored diverse modeling strategies, including mixture-of-experts architectures to handle vulnerability diversity (e.g., MoEVD~\cite{yang2025one}) and graph-based representations that capture program structure via message passing (e.g., SnapVul~\cite{wu2023learning}). 
In parallel, a large body of work leverages richer program representations, such as AST/CFG/DFG-based graphs or learned code embeddings to improve semantic modeling for vulnerability detection, but these approaches still predominantly operate within a single-function input setting~\cite{cui2020vuldetector,dong2023sedsvd,tian2024enhancing}.
Recent work such as LLMxCPG~\cite{lekssays2025llmxcpg} follows the same paradigm by constructing CPGs at the function level and using LLMs for code slicing and vulnerability detection, without incorporating broader inter-procedural context.
Our work complements these works by targeting a different bottleneck: enabling models to effectively leverage inter-procedural context under practical input constraints, through context profiling/selection and structured reasoning.

\paragraph{Vulnerability datasets}
Early commit-mined datasets such as BigVul~\cite{fan2020ac} are known to contain substantial noise~\cite{ding2024vulnerability,risse2023limits,risse2025top}, motivating a wave of higher-quality benchmarks. PrimeVul~\cite{ding2024vulnerability} improves label quality by restricting to single-function vulnerability-fixing commits and reports 86\% validity based on manual inspection.
CleanVul~\cite{li2024cleanvul} further reduces noise using LLM-based filtering and reports 91\% validity, while TitanVul~\cite{li2025out} aggregates multiple sources and applies multi-agent verification, reporting 92\% validity. 
Building on these efforts, we construct context-enriched variants of \primevul, \titanvul, and \cleanvul that preserve their high-quality labels while adding repository-derived inter-procedural context, enabling rigorous evaluation of context-aware and reasoning-based detection beyond function-only inputs.

\vspace{-2mm}
\section{Conclusion and Future Work}
\label{sec:conclusion}

Even as vulnerability datasets become cleaner and more carefully curated, most existing approaches still formulate vulnerability detection as a function-level classification problem, despite the fact that many real-world vulnerabilities depend on inter-procedural semantics beyond the target function.
Our study shows that simply appending context is not a reliable solution and can degrade strong code models.
We present \tool, a two-phase framework that (i) profiles and ranks inter-procedural context to reduce redundancy and emphasize security-relevant signals, and (ii) trains LLMs to perform structured vulnerability-oriented reasoning over the target function, context, and auxiliary metadata.
Because all three datasets are balanced, we use \emph{accuracy} as the primary metric throughout the paper.
Across \primevul, \titanvul, and \cleanvul, \tool improves accuracy over the strongest function-only baseline (UniXcoder) by +11.13 points (56.65\% $\rightarrow$ 67.78\%) on \primevul, +10.08 points (63.68\% $\rightarrow$ 73.76\%) on \titanvul, and +6.70 points (58.24\% $\rightarrow$ 64.94\%) on \cleanvul.
On \primevul, \tool also exceeds the best previously reported accuracy from recent top-venue approaches (e.g., USENIX Security'25, ICSE'25, and ICSE'26) from 55.17\% to 67.78\%.
We additionally release context-enriched variants of \primevul, \titanvul, and \cleanvul to support future work on inter-procedural vulnerability detection.
For future work, we plan to incorporate more path- and temporal-sensitive signals to better handle lifetime weaknesses such as double-free.

\cleardoublepage
\bibliographystyle{plainurl}
\bibliography{bibliography}

\end{document}

%% file: tables/motivation.tex
\begin{table}[htb]
\centering
\caption{Vulnerability detection performance on the \primevul, \titanvul, and \cleanvul datasets, comparing function-only inputs with raw context augmentation for CodeBERT and UniXcoder.}
\label{tb:empirical}
\resizebox{0.97\columnwidth}{!}{
\begin{tabular}{@{}lccccccc@{}}
\toprule
\textbf{Dataset} & \textbf{Model} & \textbf{Input} & \textbf{Acc \%} & \textbf{P \%} & \textbf{R \%} & \textbf{F1 \%} \\
\midrule
\textbf{\primevul} & \multirow{2}{*}{\shortstack{CodeBERT}} & Function Body & \textbf{53.21} & 52.78 & 63.03 & 55.62 \\
 & & w/ Raw Context & 52.51 & 52.20 & 72.90 & 59.77 \\
\cmidrule{2-7}
 & \multirow{2}{*}{\shortstack{UniXcoder}} & Function Body & \textbf{56.65} & 55.34 & 69.45 & 61.55 \\
 & & w/ Raw Context & 54.22 & 54.09 & 56.16 & 54.89 \\
\midrule
\textbf{\titanvul} & \multirow{2}{*}{\shortstack{CodeBERT}} & Function Body & \textbf{54.41} & 54.78 & 59.91 & 54.91 \\
 & & w/ Raw Context & 53.92 & 54.22 & 71.74 & 59.02 \\
\cmidrule{2-7}
 & \multirow{2}{*}{\shortstack{UniXcoder}} & Function Body & \textbf{63.68} & 61.99 & 71.10 & 66.10 \\
 & & w/ Raw Context & 62.87 & 61.67 & 68.27 & 64.74 \\
\midrule
\textbf{\cleanvul} & \multirow{2}{*}{\shortstack{CodeBERT}} & Function Body & \textbf{55.67} & 54.73 & 67.74 & 59.94 \\
 & & w/ Raw Context & 52.47 & 43.51 & 53.62 & 47.69 \\
\cmidrule{2-7}
 & \multirow{2}{*}{\shortstack{UniXcoder}} & Function Body & \textbf{58.24} & 56.62 & 70.72 & 62.84 \\
 & & w/ Raw Context & 55.22 & 54.59 & 62.28 & 57.99 \\
\bottomrule
\end{tabular}
}
\end{table}
\vspace{-2mm}

%% file: figures/overview.tex
\begin{figure*}[t]
  \centering
  \includegraphics[trim={0cm 0cm 0cm 0cm}, clip, width=\linewidth]{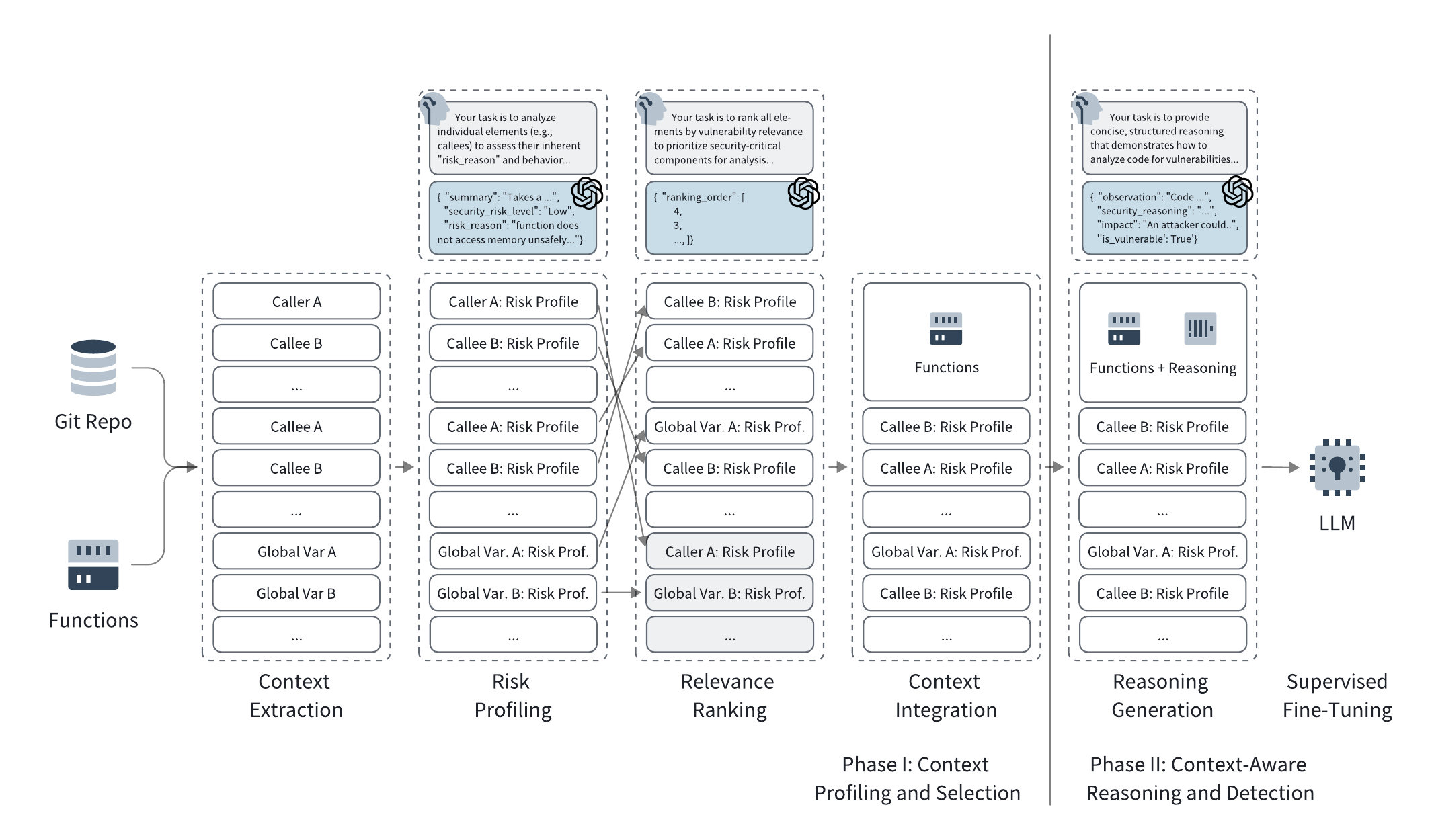}
  \caption{Overview of our context-aware vulnerability detection framework \tool. The approach operates in two phases. In Phase I, we prepare and prioritize inter-procedural context through four stages: (1) extracting callers, callees, and global variables related to a target function via CPG-based static analysis; (2) constructing security risk profiles for each contextual element; (3) ranking elements by vulnerability relevance; and (4) integrating the highest-ranked context that fits within the LLM's context window. In Phase II, the target function body, selected contextual elements (callers, callees, and globals), and auxiliary vulnerability metadata (e.g., commit messages, CVE and CWE information) are jointly provided to the LLM to generate structured reasoning traces. These reasoning traces are then used for supervised fine-tuning, training the LLM to perform vulnerability detection as a reasoning task.
  }
  \label{f:overview}
\end{figure*}
\vspace{-3mm}

%% file: tables/prompt.tex
\begin{table}[!t]
\centering
\scriptsize
\caption{Prompt templates for generating structured reasoning traces. The templates distinguish between vulnerable and non-vulnerable classes to generate reasoning traces that supervise the fine-tuning of the detection model.}
\label{tb:prompts}
\begin{tabular}{p{0.95\columnwidth}}
\toprule
\multicolumn{1}{c}{\footnotesize{Prompt for Vulnerable Functions}} \\
\midrule
You are an expert security analyst training an AI system to detect vulnerabilities in code. Your task is to provide concise, structured reasoning that demonstrates how to analyze code for security vulnerabilities.
\\
\\
You are provided with:\\
- \textbf{Code Diff}: Shows vulnerable code (before) and fixed code (after).\\
- \textbf{CVE Description}: Detailed description of the vulnerability.\\
- \textbf{CWE Information}: The vulnerability classification.\\
- \textbf{Commit Message}: Context about the security fix.\\
- \textbf{Additional Context}: Extra contextual information about the vulnerability.\\
Focus your analysis on the vulnerable code (the before version). You need to identify this code as vulnerable and provide reasoning that demonstrates why it contains security flaws. Your analysis should help train models to recognize similar vulnerable patterns.
\\
\\
Respond with only the JSON below.\\
\{\\
  "observation": "Brief description of what the code does that could be problematic",\\
  "security\_reasoning": "Explanation of why this creates a security vulnerability",\\
  "impact": "What an attacker could achieve by exploiting this vulnerability",\\
  "is\_vulnerable": true,\\
  "confidence\_score": 0--10\\
\}\\
\midrule
\multicolumn{1}{c}{\footnotesize{Prompt for Non-Vulnerable Functions}} \\
\midrule
You are an expert security analyst training an AI system to recognize safe and secure code patterns. Your task is to provide concise, structured reasoning that demonstrates how to analyze code for security safeguards using counter-factual analysis.
\\
\\
You are provided with:\\
- \textbf{Code Diff}: Shows code changes from vulnerable (before) to secure (after) implementation.\\
- \textbf{CVE Description}: Detailed description of the vulnerability that was fixed.\\
- \textbf{CWE Information}: Information about the specific security vulnerability type that was addressed.\\
- \textbf{Commit Message}: Commit message about the security fix.\\
- \textbf{Additional Context}: Extra contextual information about the vulnerability.\\
Focus your analysis on the secure/fixed code (the after version). You need to identify this code as safe and provide reasoning that demonstrates the security protections in place. Your analysis should help train models to recognize secure coding patterns.
\\
\\
Respond with only the JSON below.\\
\{\\
   "observation": "Brief description of what the code does and how it handles potential security concerns",\\
   "security\_reasoning": "Explanation of the security measures and protections in place. Include counterfactual reasoning by explaining what vulnerabilities could occur if the implementation lacked these security measures, using the vulnerable pattern from the code diff as reference",\\
   "impact": "How these security measures prevent potential attacks",\\
   "is\_vulnerable": false,\\
   "confidence\_score": 0--10\\
\}\\
\bottomrule
\end{tabular}
\end{table}
\vspace{-3mm}

%% file: tables/rq1_performance.tex
\begin{table}[htb]
\centering
\caption{Vulnerability detection performance on the \titanvul, \primevul, and \cleanvul datasets, comparing baseline methods with \tool.}
\label{tb:performance}
\resizebox{0.97\columnwidth}{!}{
\begin{tabular}{@{}lccccccc@{}}
\toprule
\textbf{Data} & \textbf{Model} & \textbf{Approach} & \textbf{Acc \%} & \textbf{P \%} & \textbf{R \%} & \textbf{F1 \%} \\
\midrule
\multirow{6}{*}{\rotatebox{90}{\textbf{\primevul}}} & \multirow{1}{*}{GPT-4.1} & Baseline & 51.90 & 51.17 & 83.30 & 63.39 \\
\cmidrule{2-7}
 & \multirow{1}{*}{\shortstack{CodeBERT}} & Baseline & 53.21 & 52.78 & 63.03 & 55.62 \\
\cmidrule{2-7}
 & \multirow{1}{*}{\shortstack{UniXcoder}} & Baseline & 56.65 & 55.34 & 69.45 & 61.55 \\
\cmidrule{2-7}
 & \multirow{1}{*}{\shortstack{Qwen2.5-7B}} & \tool & 62.70 & 59.97 & 76.40 & 67.19 \\
\cmidrule{2-7}
 & \multirow{1}{*}{\shortstack{Qwen2.5-32B}} & \tool & \textbf{67.78} & 67.62 & 68.23 & 67.93 \\
\midrule
\multirow{6}{*}{\rotatebox{90}{\textbf{\titanvul}}} & \multirow{1}{*}{GPT-4.1} & Baseline & 51.36 & 50.90 & 76.74 & 61.21 \\
\cmidrule{2-7}
 & \multirow{1}{*}{\shortstack{CodeBERT}} & Baseline & 54.41 & 54.78 & 59.91 & 54.91 \\
\cmidrule{2-7}
 & \multirow{1}{*}{\shortstack{UniXcoder}} & Baseline & 63.68 & 61.99 & 71.10 & 66.10 \\
\cmidrule{2-7}
 & \multirow{1}{*}{\shortstack{Qwen2.5-7B}} & \tool & 68.46 & 69.23 & 66.46 & 67.81 \\
\cmidrule{2-7}
 & \multirow{1}{*}{\shortstack{Qwen2.5-32B}} & \tool & \textbf{73.76} & 72.88 & 76.27 & 74.54 \\
\midrule
\multirow{6}{*}{\rotatebox{90}{\textbf{\cleanvul}}} & \multirow{1}{*}{GPT-4.1} & Baseline & 51.83 & 51.21 & 77.24 & 61.59 \\
\cmidrule{2-7}
 & \multirow{1}{*}{\shortstack{CodeBERT}} & Baseline & 55.67 & 54.73 & 67.74 & 59.94 \\
\cmidrule{2-7}
 & \multirow{1}{*}{\shortstack{UniXcoder}} & Baseline & 58.24 & 56.62 & 70.72 & 62.84 \\
\cmidrule{2-7}
 & \multirow{1}{*}{\shortstack{Qwen2.5-7B}} & \tool & 62.18 & 64.32 & 54.71 & 59.13 \\
\cmidrule{2-7}
 & \multirow{1}{*}{\shortstack{Qwen2.5-32B}} & \tool & \textbf{64.94} & 66.00 & 61.60 & 63.73 \\
\bottomrule
\end{tabular}
}
\end{table}
\vspace{-3mm}

%% file: tables/rq1_sota.tex
\begin{table}[htb]
\centering
\caption{Vulnerability detection performance on the \primevul dataset, comparing \tool with representative state-of-the-art methods.}
\label{tb:sota}
\resizebox{0.8\columnwidth}{!}{
\begin{tabular}{@{}lcccc@{}}
\toprule
\textbf{Approach} & \textbf{Acc \%} & \textbf{P \%} & \textbf{R \%} & \textbf{F1 \%} \\
\midrule
LLMxCPG \cite{lekssays2025llmxcpg} & 50.36 & 50.51 & 35.75 & 41.87 \\
Ding et al.'s CoT \cite{ding2024vulnerability} & 51.26 & 53.55 & 19.08 & 28.14 \\
GPTLens \cite{hu2023large} & 51.84 & 51.44 & 65.52 & 57.63 \\
VulTrial \cite{widyasari2025let} & 55.17 & 54.95 & 57.47 & 56.18 \\
\midrule
\tool & \textbf{67.78} & 67.62 & 68.23 & 67.93 \\
\bottomrule
\end{tabular}
}
\end{table}
\vspace{-3mm}

%% file: tables/rq2_cwe.tex
\begin{table*}[htb]
\centering
\caption{Per-CWE performance for the 15 most frequent vulnerability types in \primevul and \titanvul, comparing \tool (Qwen2.5-Coder-32B) against the UniXcoder function-only baseline.}
\label{tb:cwe}
\resizebox{0.8\textwidth}{!}{
\begin{tabular}{@{}llccccccccccc@{}}
\toprule
\multirow{2}{*}{\textbf{Dataset}} & \multirow{2}{*}{\textbf{CWE}} & \multirow{2}{*}{\textbf{Count}} & \multicolumn{4}{c}{\textbf{Baseline}} & \multicolumn{4}{c}{\textbf{\tool}} & \multicolumn{2}{c}{\textbf{Improvement}} \\
\cmidrule(lr){4-7} \cmidrule(lr){8-11} \cmidrule(lr){12-13}
& & & \textbf{Acc \%} & \textbf{P \%} & \textbf{R \%} & \textbf{F1 \%} & \textbf{Acc \%} & \textbf{P \%} & \textbf{R \%} & \textbf{F1 \%} & \textbf{$\Delta$ Acc} & \textbf{$\Delta$ F1} \\
\midrule
\multirow{1}{*}{\textbf{\primevul}} & 119 & 124 & 60.48 & 60.00 & 62.90 & 61.42 & \textbf{77.42} & 73.61 & 85.48 & 79.10 & +16.94 & +17.68 \\
 & 125 & 116 & 62.07 & 60.29 & 70.69 & 65.08 & \textbf{67.24} & 70.00 & 60.34 & 64.81 & +5.17 & -0.27 \\
 & 20 & 92 & 59.78 & 57.63 & 73.91 & 64.76 & \textbf{66.30} & 68.29 & 60.87 & 64.37 & +6.52 & -0.39 \\
 & 787 & 86 & 59.30 & 58.33 & 65.12 & 61.54 & \textbf{61.63} & 60.00 & 69.77 & 64.52 & +2.33 & +2.98 \\
 & 416 & 58 & 46.55 & 47.37 & 62.07 & 53.73 & \textbf{53.45} & 53.33 & 55.17 & 54.24 & +6.90 & +0.51 \\
 & 476 & 56 & 55.36 & 53.85 & 75.00 & 62.69 & \textbf{69.64} & 65.71 & 82.14 & 73.02 & +14.28 & +10.33 \\
 & 200 & 56 & 53.57 & 52.78 & 67.86 & 59.37 & \textbf{76.79} & 80.00 & 71.43 & 75.47 & +23.22 & +16.10 \\
 & 703 & 52 & 55.77 & 54.29 & 73.08 & 62.30 & \textbf{69.23} & 65.62 & 80.77 & 72.41 & +13.46 & +10.11 \\
 & 264 & 48 & 52.08 & 51.52 & 70.83 & 59.65 & \textbf{79.17} & 79.17 & 79.17 & 79.17 & +27.09 & +19.52 \\
 & 190 & 42 & 64.29 & 65.00 & 61.90 & 63.41 & \textbf{69.05} & 68.18 & 71.43 & 69.77 & +4.76 & +6.36 \\
 & 399 & 40 & 50.00 & 50.00 & 75.00 & 60.00 & \textbf{70.00} & 68.18 & 75.00 & 71.43 & +20.00 & +11.43 \\
 & 189 & 40 & 55.00 & 55.56 & 50.00 & 52.63 & \textbf{70.00} & 83.33 & 50.00 & 62.50 & +15.00 & +9.87 \\
 & 369 & 26 & 65.38 & 66.67 & 61.54 & 64.00 & \textbf{73.08} & 68.75 & 84.62 & 75.86 & +7.70 & +11.86 \\
 & 415 & 24 & 50.00 & 50.00 & 83.33 & 62.50 & \textbf{62.50} & 71.43 & 41.67 & 52.63 & +12.50 & -9.87 \\
 & 401 & 20 & 50.00 & 50.00 & 90.00 & 64.29 & \textbf{65.00} & 66.67 & 60.00 & 63.16 & +15.00 & -1.13 \\
 \cmidrule{2-13}
 & Avg. & -- & -- & -- & -- & -- & -- & -- & -- & -- & +12.72 & +7.01 \\
\midrule
\multirow{1}{*}{\textbf{\titanvul}} & 125 & 243 & 77.37 & 76.61 & 78.51 & 77.55 & \textbf{84.36} & 84.87 & 83.47 & 84.17 & +6.99 & +6.62 \\
 & 787 & 226 & 78.32 & 73.68 & 87.50 & 80.00 & \textbf{81.86} & 83.33 & 81.08 & 82.19 & +3.54 & +2.19 \\
 & 119 & 214 & 57.48 & 55.63 & 73.83 & 63.45 & \textbf{74.77} & 73.68 & 78.50 & 76.02 & +17.29 & +12.57 \\
 & 20 & 116 & 59.48 & 57.53 & 72.41 & 64.12 & \textbf{69.83} & 70.18 & 68.97 & 69.57 & +10.35 & +5.45 \\
 & 416 & 84 & 59.52 & 58.00 & 69.05 & 63.04 & \textbf{66.28} & 66.67 & 65.12 & 65.88 & +6.76 & +2.84 \\
 & 476 & 70 & 58.57 & 57.14 & 68.57 & 62.34 & \textbf{73.61} & 74.29 & 72.22 & 73.24 & +15.04 & +10.90 \\
 & 264 & 66 & 68.18 & 65.79 & 75.76 & 70.42 & \textbf{85.14} & 86.11 & 86.11 & 86.11 & +16.96 & +15.69 \\
 & 200 & 66 & 62.12 & 59.09 & 78.79 & 67.53 & \textbf{71.21} & 66.67 & 84.85 & 74.67 & +9.09 & +7.14 \\
 & 190 & 66 & 71.21 & 66.67 & 84.85 & 74.67 & \textbf{78.79} & 75.68 & 84.85 & 80.00 & +7.58 & +5.33 \\
 & 362 & 66 & 68.18 & 66.67 & 72.73 & 69.57 & \textbf{75.00} & 72.97 & 79.41 & 76.06 & +6.82 & +6.49 \\
 & 399 & 62 & 56.45 & 55.26 & 67.74 & 60.87 & \textbf{71.88} & 69.44 & 80.65 & 74.63 & +15.43 & +13.76 \\
 & 120 & 44 & 75.00 & 73.91 & 77.27 & 75.56 & \textbf{93.18} & 91.30 & 95.45 & 93.33 & +18.18 & +17.77 \\
 & 189 & 40 & \textbf{62.50} & 61.90 & 65.00 & 63.41 & 61.90 & 66.67 & 60.00 & 63.16 & -0.60 & -0.25 \\
 & 269 & 32 & 71.88 & 68.42 & 81.25 & 74.29 & \textbf{78.12} & 76.47 & 81.25 & 78.79 & +6.24 & +4.50 \\
 & 415 & 29 & \textbf{68.97} & 68.75 & 73.33 & 70.97 & 62.07 & 66.67 & 53.33 & 59.26 & -6.90 & -11.71 \\
 \cmidrule{2-13}
 & Avg. & -- & -- & -- & -- & -- & -- & -- & -- & -- & +8.85 & +6.62 \\
\bottomrule
\end{tabular}
}
\end{table*}
\vspace{-3mm}

%% file: tables/rq3.tex
\begin{table}[t]
\centering
\caption{Evaluation of context integration strategies using Qwen2.5-Coder-32B.}
\label{tb:integration}
\resizebox{0.48\textwidth}{!}{
\begin{tabular}{@{}lcccccc@{}}
\toprule
\textbf{Dataset} & \textbf{Integration Strategy} & \textbf{Acc \%} & \textbf{P \%} & \textbf{R \%} & \textbf{F1 \%} \\
\midrule
\textbf{\primevul} & Insert-Before & 65.60 & 64.01 & 71.68 & 67.63 \\
 & Insert-After & \textbf{67.78} & 67.62 & 68.23 & 67.93 \\
 & As-Comments & 64.60 & 63.66 & 68.05 & 65.78 \\
\midrule
\textbf{\titanvul} & Insert-Before & 72.00 & 71.56 & 73.55 & 72.54 \\
 & Insert-After & \textbf{73.76} & 72.88 & 76.27 & 74.54 \\
 & As-Comments & 72.26 & 71.52 & 74.05 & 72.76 \\
\midrule
\textbf{\cleanvul} & Insert-Before & 62.41 & 63.04 & 60.00 & 61.84 \\
 & Insert-After & \textbf{64.94} & 66.00 & 61.60 & 63.73 \\
 & As-Comments & 63.44 & 61.91 & 69.88 & 65.65 \\
\bottomrule
\end{tabular}
}
\end{table}
\vspace{-3mm}

%% file: tables/rq4_ab1.tex
\begin{table}[t]
\centering
\caption{Ablation study results for model variants with different settings using Qwen2.5-Coder-32B.} 
\label{tb:ablation}
\resizebox{0.97\columnwidth}{!}{
\begin{tabular}{@{}llcccccc@{}}
\toprule
\textbf{Data} & \textbf{Approach} & \textbf{Acc \%} & \textbf{P \%} & \textbf{R \%} & \textbf{F1 \%} & \textbf{$\Delta$ Acc (pts)} \\
\midrule
\textbf{\primevul} & \tool & \textbf{67.78} & 67.62 & 68.23 & 67.93 & -- \\
\cmidrule{2-7}
 & w/o Phase I & 65.42 & 62.87 & 75.31 & 68.53 & -2.36 \\
\cmidrule{2-7}
 & w/o Phase II & 52.54 & 52.11 & 69.58 & 59.59 & -15.24 \\
\midrule
\textbf{\titanvul} & \tool & \textbf{73.76} & 72.88 & 76.27 & 74.54 & -- \\
\cmidrule{2-7}
 & w/o Phase I & 72.20 & 70.76 & 75.66 & 73.13 & -1.56 \\
\cmidrule{2-7}
 & w/o Phase II & 52.38 & 54.33 & 31.46 & 39.84 & -21.38 \\
\midrule
\textbf{\cleanvul} & \tool & \textbf{64.94} & 66.00 & 61.60 & 63.73 & -- \\
\cmidrule{2-7}
 & w/o Phase I & 62.52 & 62.03 & 64.59 & 63.28 & -2.42 \\
\cmidrule{2-7}
 & w/o Phase II & 52.41 & 55.68 & 23.68 & 33.23 & -12.53 \\
\bottomrule
\end{tabular}
}
\end{table}
\vspace{-3mm}

%% file: tables/rq4_ab2.tex
\begin{table}[t]
\centering
\caption{Vulnerability detection performance comparing function-body-only inputs with adding \tool's Phase I processed context.}
\label{tb:empirical_2}
\resizebox{0.97\columnwidth}{!}{
\begin{tabular}{@{}lccccccc@{}}
\toprule
\textbf{Dataset} & \textbf{Model} & \textbf{Input} & \textbf{Acc \%} & \textbf{P \%} & \textbf{R \%} & \textbf{F1 \%} \\
\midrule
\textbf{\primevul} & \multirow{2}{*}{\shortstack{CodeBERT}} & Function Body & \textbf{53.21} & 52.78 & 63.03 & 55.62 \\
 & & w/ Phase I & 52.11 & 53.99 & 27.22 & 33.37 \\
\cmidrule{2-7}
 & \multirow{2}{*}{\shortstack{UniXcoder}} & Function Body & \textbf{56.65} & 55.34 & 69.45 & 61.55 \\
 & & w/ Phase I & 55.26 & 54.42 & 64.73 & 59.05 \\
\midrule
\textbf{\titanvul} & \multirow{2}{*}{\shortstack{CodeBERT}} & Function Body & \textbf{54.41} & 54.78 & 59.91 & 54.91 \\
 & & w/ Phase I & 52.86 & 36.44 & 31.77 & 33.82 \\
\cmidrule{2-7}
 & \multirow{2}{*}{\shortstack{UniXcoder}} & Function Body & \textbf{63.68} & 61.99 & 71.10 & 66.10 \\
 & & w/ Phase I & 62.88 & 62.18 & 65.89 & 63.93 \\
\midrule
\textbf{\cleanvul} & \multirow{2}{*}{\shortstack{CodeBERT}} & Function Body & \textbf{55.67} & 54.73 & 67.74 & 59.94 \\
 & & w/ Phase I & 52.60 & 54.36 & 50.65 & 45.62 \\
\cmidrule{2-7}
 & \multirow{2}{*}{\shortstack{UniXcoder}} & Function Body & \textbf{58.24} & 56.62 & 70.72 & 62.84 \\
 & & w/ Phase I & 54.75 & 55.16 & 51.41 & 52.49 \\
\bottomrule
\end{tabular}
}
\end{table}
\vspace{-2mm}